\documentclass[prl, reprint, superscriptaddress, showkeys]{revtex4-2}

\usepackage{graphicx}
\usepackage{dcolumn}
\usepackage{bm}
\usepackage{hyperref}
\usepackage{xcolor}
\usepackage{caption}
\usepackage{subcaption}
\usepackage{float}
\usepackage{textcomp}
\usepackage{physics}

\begin{document}

\date{\today}


\author{Aitor \surname{De Andres}}
\email[e-mail:]{aitor.de.andres@umu.se}
\affiliation{Department of Physics, Umeå University, Linnaeus väg 24, 90187, Umeå, Sweden}
\author{Shikha \surname{Bhadoria}}
\email[e-mail:]{shikha.bhadoria@physics.gu.se}
\author{Javier Tello \surname{Marmolejo}}
\affiliation{Department of Physics, University of Gothenburg, Origovägen 6B 41296 Göteborg, Sweden}
\author{Alexander \surname{Muschet}}
\author{Peter \surname{Fischer}}
\altaffiliation[Present address: ]
{Marvel Fusion GmbH, Theresienhöhe 12, 80339 Munich, Germany}
\author{Hamid \surname{Reza Barzegar}}
\affiliation{Department of Physics, Umeå University, Linnaeus väg 24, 90187, Umeå, Sweden}
\author{Tom \surname{Blackburn}}
\affiliation{Department of Physics, University of Gothenburg, Origovägen 6B 41296 Göteborg, Sweden}

\author{Arkady \surname{Gonoskov}}
\affiliation{Department of Physics, University of Gothenburg, Origovägen 6B 41296 Göteborg, Sweden}
\author{Dag \surname{Hanstorp}}
\affiliation{Department of Physics, University of Gothenburg, Origovägen 6B 41296 Göteborg, Sweden}
\author{Mattias \surname{Marklund}}
\affiliation{Department of Physics, University of Gothenburg, Origovägen 6B 41296 Göteborg, Sweden}
\author{Laszlo \surname{Veisz}}
\email[e-mail:]{laszlo.veisz@umu.se}
\affiliation{Department of Physics, Umeå University, Linnaeus väg 24, 90187, Umeå, Sweden}

\preprint{APS/123-QED}

\begin{abstract}
Acceleration of electrons in vacuum directly by intense laser fields, often termed vacuum laser acceleration (VLA), holds great promise for the creation of compact sources of high-charge, ultrashort, relativistic electron bunches. However, while the energy gain is expected to be higher with tighter focusing (i.e. stronger electric field), this does not account for the reduced acceleration range, which is limited by diffraction.
Here, we present the results of an experimental investigation of VLA, using tungsten nanotips driven by relativistic-intensity few-cycle laser pulses. We demonstrate the acceleration of relativistic electron beams with typical charge of 100s pC to 15 MeV energies. Two different focusing geometries (tight and loose, with f-numbers one and three respectively) produced comparable results, despite a factor of ten difference in the peak intensities, which is evidence for the importance of post-injection acceleration mechanisms around the focus. Our results are in good agreement with the results of full-scale, three-dimensional particle-in-cell simulations.

\end{abstract}

\title{Unforeseen advantage of looser focusing in vacuum laser acceleration}

\maketitle

Vacuum laser acceleration (VLA) is a particle acceleration paradigm where electrons gain net energy from the interaction with a laser field in vacuum. The maximum accelerating electric field in VLA, which is controlled by the laser power and focusing geometry, exceeds $1$~TV/m for typical state-of-the-art multi-TW lasers. Thus, it significantly exceeds the field strength in other acceleration scenarios, such as conventional radio-frequency acceleration \cite{Humphries2013}, dielectric laser acceleration \cite{Peralta2013}, direct laser acceleration \cite{gahn1999multi}, laser-wakefield acceleration or plasma-wakefield acceleration \cite{Esarey2009}. This makes VLA a promising candidate for a future electron source due to its high accelerating field, short acceleration length and short bunch duration. 

However, the interplay between the physical mechanisms that drive VLA in the relativistic regime is not fully understood. Two relevant acceleration mechanisms have been identified in simulations: \textit{capture and acceleration} \cite{wang2001vacuum, pang2002subluminous}, taking place over many Rayleigh ranges with moderate accelerating gradients, and \textit{focal spot acceleration} \cite{popov2008electron}, concentrated within one Rayleigh range around the focus with much stronger gradients. However, none of the theoretical models \cite{Esarey1995, Hartemann1995, Cheng1999, Salamin.2002, hu2002, popov2008electron, Ramsey2020} have been experimentally verified. This is due to the challenging requirements of VLA. The initial electron bunches must have (1) relativistic energy, such that they propagate with the laser for a certain distance and (2) sub-femtosecond duration, such that they fit in the half-cycle long accelerating phase of the electromagnetic pulse. 

There are only a few (sometimes debated) experiments in the ponderomotive regime \cite{Malka1997, mora1998comment, mcdonald1998comment, Plettner2005, Cline2013, Braenzel2017}, which report non- or slightly relativistic energies. Alternatively, radially polarized laser pulses were also proposed for VLA \cite{varin2006, wong2017laser}, but only non-relativistic energies have been demonstrated \cite{payeur2012, carbajo2016}.
Recent works have realized VLA with linear laser polarization up to multi-MeV energies from different objects as an electron source, such as large fused silica targets \cite{Thevenet2016}, nanotips \cite{Cardenas2019}, and thin foils \cite{Singh2022}. However, neither the underlying celeration mechanism has been identified nor the competing roles of electric field strength and acceleration distance have been investigated.

Here, we report the result of an experimental campaign using the sub-5-fs Light Wave Synthesizer 20 \cite{Rivas2017} to drive VLA from nanotips, using two different focusing geometries, as characterized by their differing f-numbers (f\#). We investigate for the first time the dependence of the VLA process on the focusing geometry and how electrons dephase in the accelerating laser field.
We show that comparable electron energies (around 15 MeV) are obtained in both loose and tight focusing despite an order of magnitude difference in laser intensity. We explain this in terms of the interplay between laser electric field strength and accelerating distance. We also measure a hole in the electron angular distribution, which is caused by the influence of the laser and provides further evidence that VLA is the relevant process \cite{Thevenet2016, Singh2022}.

\begin{figure}
    \includegraphics[width=65mm]{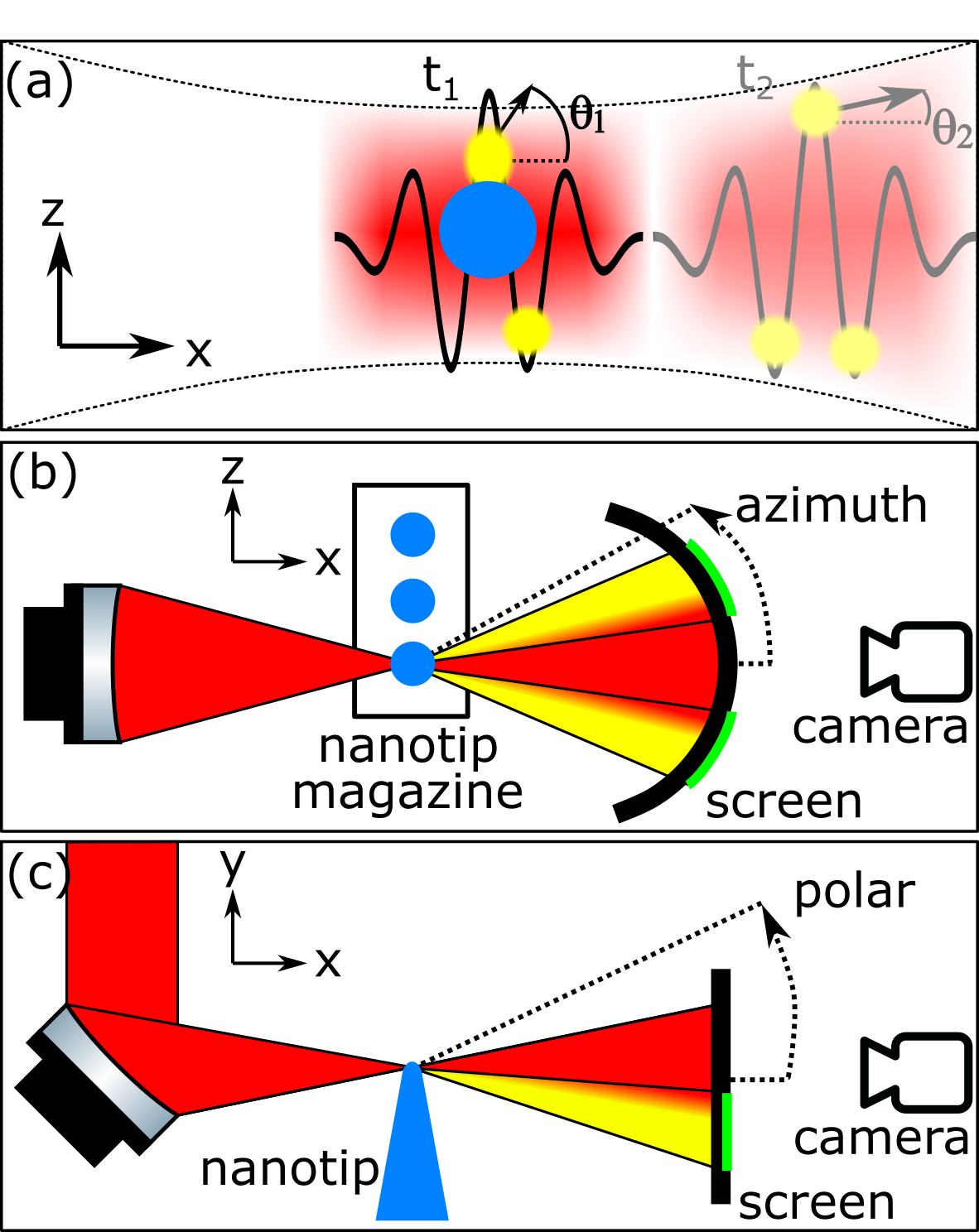}
    \caption{Acceleration mechanism in two steps: (a) electron bunches (yellow) are extracted from a nanotip (blue) at $t_1$ and accelerated via VLA around $t_2$. Experimental setup: top (b) and side (c) view. The nanotip magazine positions the tips into the laser focus (laser in red). Electrons are diagnosed by scintillating screen detectors or a spectrometer (not shown). The laser polarization direction (along $z$) is perpendicular to the axis of the nanotip.}    
    \label{fig:principle}
\end{figure}

Our experimental approach to realize VLA is to place a nanometric target, e.g. the apex of a nanotip, in the high-intensity ($I \geq 10^{18}$ W/cm\textsuperscript{2}) focus of a laser pulse \cite{popov2009vacuum, liseykina2010relativistic, andreev2013generation, naumova2004}, as demonstrated with a few-cycle laser in \cite{Cardenas2019}. In this case, the acceleration takes place in two steps \cite{Lucchio2015, Horny2021}, as illustrated in Fig.~\ref{fig:principle}(a). Initially the laser ionizes the tip into a highly overdense plasma. Then the combined laser field and the fields of the nanoplasma extract one bunch of electrons every half optical cycle, accelerating them to a relativistic initial energy at time $t_1$. Every second bunch propagates at an initial angle $\theta_1$ to the laser propagation direction as approximately predicted by the Mie theory. In the second stage, the bunch undergoes vacuum laser acceleration, becoming increasingly aligned with the laser propagation direction around time $t_2$. Finally, the acceleration terminates when the electrons escape the region of strong fields after a distance of about one Rayleigh length, with a final angle $\theta_2$. By using a sub-two cycle laser pulse with a certain carrier-envelope phase (CEP), we are able to generate in one of the two directions an essentially isolated electron bunch.

The laser pulses in the experiment were generated with the Light Wave Synthesizer 20 system \cite{Rivas2017}. Pulses with 740 nm central wavelength, 4.8 fs full width at half maximum (FWHM) duration, 70-80~mJ energy, and linear polarization perpendicular to the nanontip, were sent to the experimental chamber, where $\approx 40$~mJ was delivered to target. Two different focusing configurations were used: f\#1 and f\#3. The FWHM spot sizes were 1.23~\textmu m and 3.65~\textmu m, respectively. These correspond to peak intensities of $I_0 = 2.4 \times 10^{20}$~W/cm$^2$ and $3.6 \times 10^{19}$~W/cm$^2$, 
or to normalized vector potentials $a_0=eE_0/m_ec\omega_0$ of 9.8 and 3.8, respectively. Here $e$ is the electron charge, $m_e$ is the electron mass, $E_0$ is the laser electric field strength, $\omega_0$ the central laser frequency and $c$ the velocity of light in vacuum.
The estimated Rayleigh length (Z\textsubscript{R}) is 4.6~\textmu m with  f\#1 and 40~\textmu m with  f\#3. 
\begin{figure}
    \centering
    \includegraphics[width=75mm]{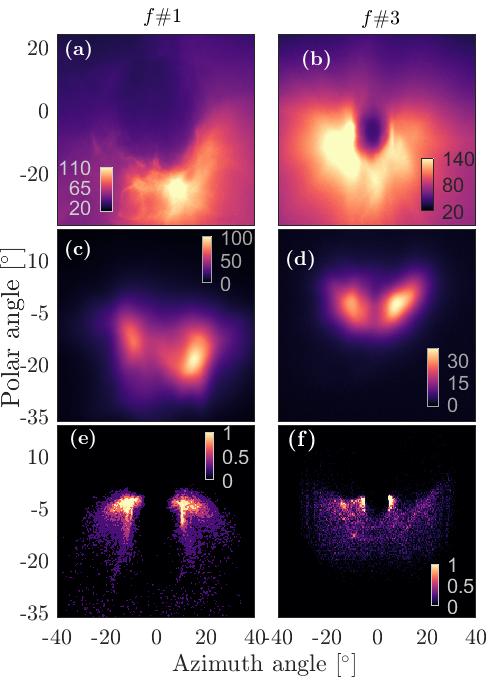}
    \caption{
    Electron angular charge distribution (pC/sr) for f\#1 (left panels) and f\#3 (right panels) focusing: all electrons above 130~keV (a, b); electrons above 1.5~MeV (c, d); simulated distributions above 0.05~MeV (e) and 0.2~MeV (f). Experimental results [(a) to (d)] are averaged over a number of shots.}
    
    \label{fig:angdist_F1}
\end{figure}
For each laser shot (whether f\#1 or f\#3), a new tungsten nanotip (apex diameter $\leq$ 100 nm) \cite{Lucier2004} was positioned with sub-\textmu m accuracy in the focus, while the laser intensity was reduced to $<10^{12}$ W/cm$^2$ (which was the measured damage threshold) to avoid premature damage to the tip. After the alignment, a single laser pulse at full power was released. The accelerated electrons were characterized by measuring their total charge and their angular distribution with an absolutely calibrated scintillating screen \cite{Buck2010, Kurz2018}, as schematically shown in Fig.~\ref{fig:principle}(b)-(c). A 1~mm thick lead shield was placed before the screen in some shots to filter out low-energy electrons ($<$1.5 MeV). 
The angle-resolved energy spectrum was also measured, using a dipole spectrometer with a scintillating screen and a CCD detector \cite{Holgreson2020}. The electron spectrometer could be rotated around the azimuth angle and had an acceptance half-angle of 7.6° and 5.5° for the f\#1 and f\#3 focusing configurations, respectively. The laser light was filtered out before reaching the detectors by placing a 10~\textmu m thin aluminum foil 
in front of the scintillating screens.

\begin{figure}
    \centering
    \includegraphics[width=85mm]{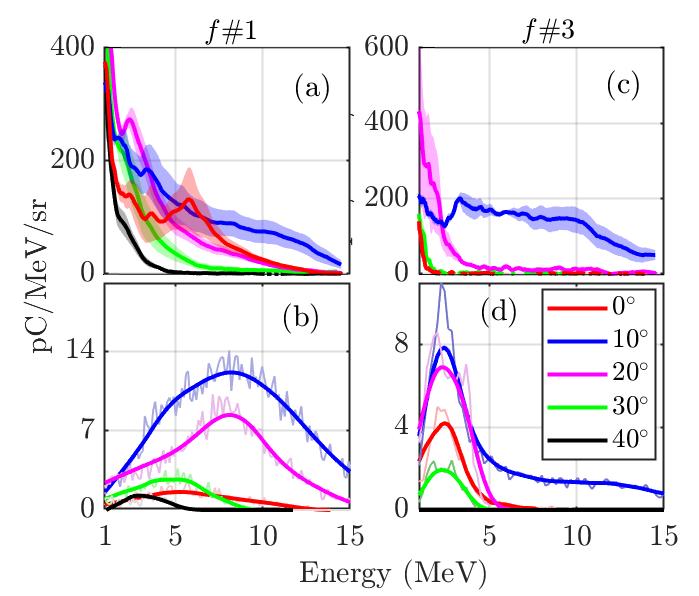}
    \caption{
    Angle-resolved electron spectra with the f$\#1$ (a,b) and f$\#3$ (c,d) configuration. (a,c) are experimental averaged electron energy spectra resolved along the azimuth angle. The shaded area in  (a,c) represents the standard deviation over several (3-5) shots. (b,d) are simulated angle-resolved electron spectra at certain azimuth angles. The lighter lines in background of (b,d) are raw simulation data and the thick lines are their smoothed version.}
    \label{fig:elspec_F1F3}
\end{figure}

The electron angular charge distributions are compared for an experiment with f\#1 and another one with f\#3 focusing in Fig.~\ref{fig:angdist_F1}. Panels (a) and (b) show distributions of electrons above 130 keV with a broad range of propagation directions, with a hole in the center for f\#3. We confirmed that the hole is related to the diffraction of the laser beam by halving the beam diameter with an iris, which reduced the diameter of the hole (not shown). Despite the difference in intensity, the overall average charge is comparable in both cases: 179 pC (432 pC in best case) and 265 pC (369 pC in best case) for f\#1 and f\#3, respectively. Inserting the Pb filter (panels (c) and (d)) shows electrons above 1.5 MeV and reveals two distinct peaks, displaced from the center along the polarization direction. The average charge is decreased to 29 pC (56 pC in best case) and 5.4~pC (12.8 pC in best case) for f\#1 and f\#3, respectively. This angular pattern was repeatedly observed despite shot-to-shot fluctuation especially in the polar angle ($\pm$10°).

Due to the reduced intensity for f\#3 focusing, the accelerated beams are expected to have lower total charge and lower electron energy. However, this is not the case, as visible from the angle-resolved energy spectra shown in Fig.~\ref{fig:elspec_F1F3}(a) and (c). For f\#1, maximal energies $> 15$~MeV were observed at 10° from the laser axis, although electrons with $> 5$~MeV were observed across the broad range of azimuthal angles (0° to 30°). For f\#3, similarly high maximal energies were observed ($> 15$~MeV), albeit in a narrower angular range: e.g. at 0° and 20° no electrons with $> 5$~MeV are seen.



\begin{figure}[htb]
\centering 
\includegraphics[width=9cm,height=3.5cm]{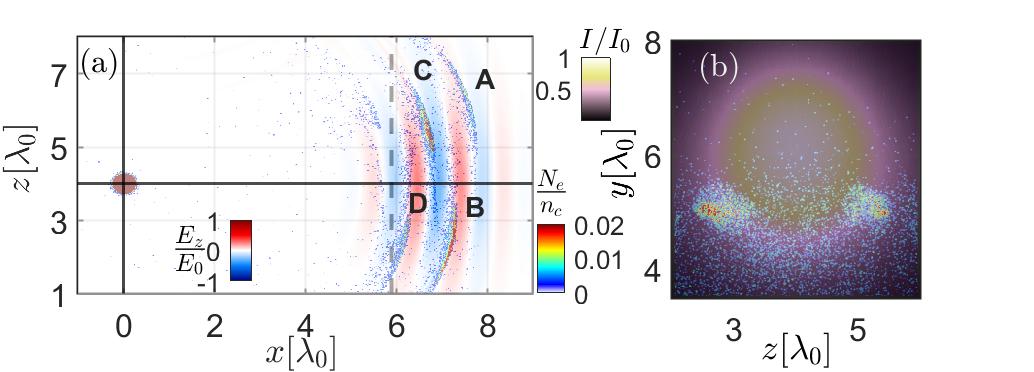}
\caption{Simulated electron distribution: (a) top view of $f\#1$ case, with laser electric field plotted over the electron density in $(x$-$z)$ space. Each individual electron bunch is labelled from A to D and the dotted gray line is one Rayleigh length from the focus at nanotip (at the center of the black solid cross). (b) Total moving electron density in $(y$-$z)$ space, plotted over the average laser intensity.}
\label{fig:PIC_F1_explanation}
\end{figure}
In order to gain insights into the experimental results, we performed particle-in-cell (PIC) simulations using Smilei-v4.7.~\cite{DEROUILLAT.2018}. By tracking particles over a sufficiently long distance we are able to explain the high performance of f\#3 focusing, as well as the angular structure of the electron beams. The simulations were performed in full 3D using a moving window technique to follows the accelerated electrons. The two focusing geometries were modelled by injecting a Gaussian laser in time (FWHM duration 4.5~fs) and in space with a focal spot size (FWHM) of 1.2~\textmu m or 3.6~\textmu m (amplitude $a_0 = 9.5$ or $a_0 = 3.7$, Rayleigh length $Z_\text{R} = 4.4$~\textmu m or 40~\textmu m) for f\#1 or f\#3, respectively. The nanotip was modelled as a structured tungsten plasma, with electron density of $100 n_c$, temperature of $5$~keV and ion charge-to-mass ratio of $44e/184m_p$. Further details may be found in Supplementary Material.

\begin{figure}[htb]
\centering 
\includegraphics[width=8.2cm]{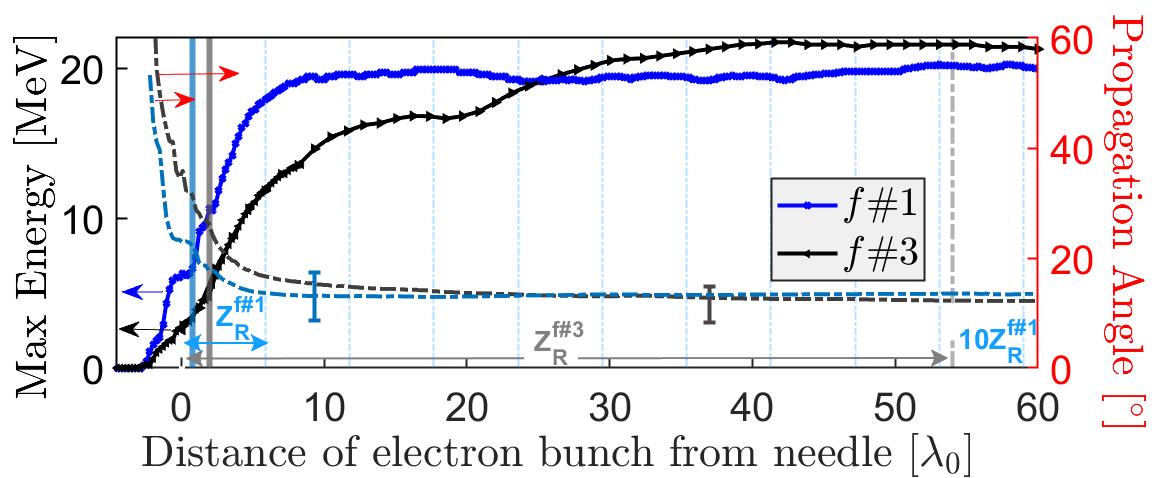}
\caption{Evolution of maximum electron energy and expected propagation angle of highest energy electrons from the plane wave estimate (see text) with distance from nanotip for $f\#1$ (blue) and $f\#3$ (black). 
The two data points are average azimuth angle from PIC simulations (error bar due to
angular spread). Dashed vertical lines indicate a Rayleigh length and solid lines discriminates VLA from nanotip field acceleration.}
\label{fig:MaxEn}
\end{figure}

We begin by showing the spatial and angular distributions of the accelerated electrons, when they have travelled approximately one Rayleigh length from the nanotip. Figure~\ref{fig:PIC_F1_explanation}~(a) (the view in the $xz$-plane) shows a train of electron bunches (labelled A to D) that are spaced half a wavelength apart and propagate with the laser pulse, with a deflection from the $x$ axis that alternates from bunch to bunch. The number density of individual electron bunches in this train depends on the CEP of the laser, as previously investigated in \cite{Cardenas2019, Horny2021}, and with a CEP of $\sim 0.3\pi$ rad, an isolated electron bunch is generated on one side. The spatial distribution in the $yz$-plane, shown in Fig.~\ref{fig:PIC_F1_explanation}~(b), reveals two distinct peaks outside the region of highest intensity, in locations that correspond to azimuth angles of about $\pm13^\circ$ and a polar angle of about $-5^\circ$, in good agreement with the experiments. The expected propagation angle for an initially resting electron in a plane wave is \cite{Gibbon2005} $\theta_2 = \arctan[\sqrt{2/(\gamma-1)}]$, where $\gamma$ is the Lorentz factor of the electrons. This predicts azimuth angles 13-$24^\circ$ for electron energies of 5-18 MeV. This is also visible in Fig.~\ref{fig:MaxEn} that plots the maximum electron energies and the corresponding calculated propagation angle of the bunches for both focusing geometries. The simulated propagation angles converge towards the theoretically predicted value by $\arctan[\sqrt{2/(\gamma-1)}]$ and are consistent with the experimental results.

The simulated and measured angular charge distributions are compared in Fig.~\ref{fig:angdist_F1}. In the simulations in Fig.~\ref{fig:angdist_F1}(e) for f\#1 the presence of two distinct peaks is reproduced well, and in Fig.~\ref{fig:angdist_F1}(f) for f\#3 the anticipated 'hole' in the distribution is also visible \cite{Thevenet2016, Singh2022}. The simulated electron energy spectra, shown in Fig.~\ref{fig:elspec_F1F3}~(b) and (d), are also in good agreement with experimental results: the highest energies ($>$15~MeV) and greatest charge is found for 10°. For f\#1 the distribution of these energetic electrons extends towards 20° and slightly towards 0°. Notably, for f\#3 a plateau extending up to about 15~MeV is clearly seen for 10°, while almost no electrons exceed 5~MeV for other angles.

\begin{figure}[htb]
\centering 
\includegraphics[width=9cm, height=7.1cm]{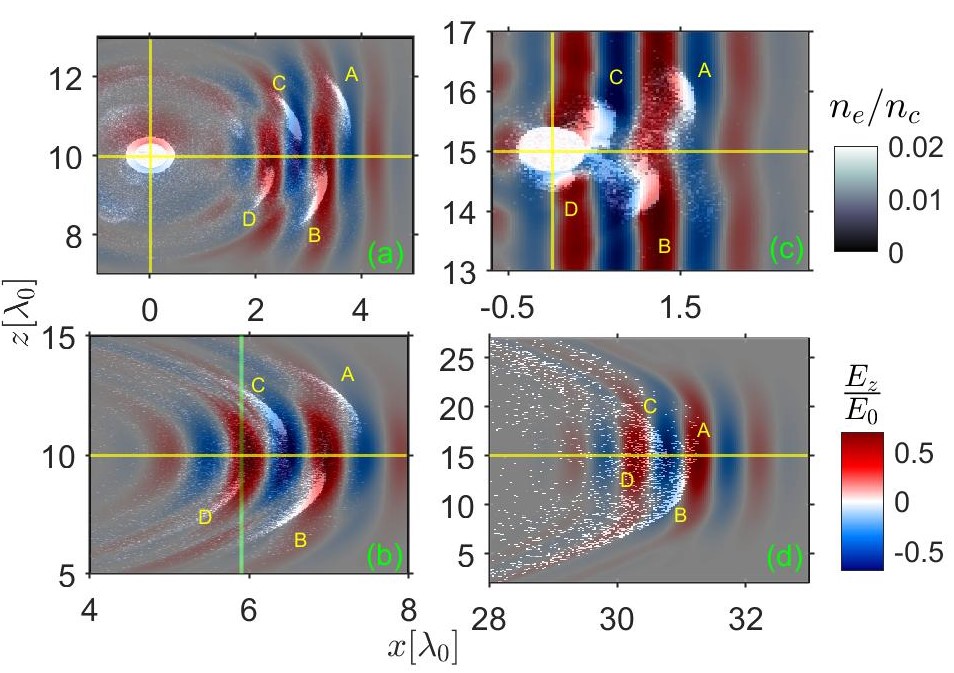}
\caption{Dephasing for both f\#1 (a, b) and f\#3 (c, d), as seen by the location and spread of the electron bunches (electron density normalized to the critical density [$n_c=2\times10^{21}\text{cm}^{-3}$] in black-white) in the laser electric field (red-blue colorscale).
The electrons have travelled (b) one or (c) half a Rayleigh length for f\#1 and f\#3, respectively.}
\label{fig:dephasing}
\end{figure}
Now, that we have confirmed that our simulations reproduce the physics of experiment, we discuss why the weaker electromagnetic field provided by f\#3 focusing yields the same maximal electron energies. The evolution of the electron bunch energy as a function of distance from the nanotip is shown in Fig.~\ref{fig:MaxEn}. For f\#1 the energy saturates at about one Rayleigh length \cite{Cardenas2019, popov2008electron} and does not change up to ten Rayleigh lengths, where the laser intensity is lower than the relativistic limit. Therefore, we conclude that the focal spot acceleration mechanism is relevant for our case and electrons are accelerated up to the Rayleigh range by a strong and almost constant electric field. The electron energy may be estimated according to some basic considerations. The electron energy change from the injection point ($t_1$ and $x_1$) is given by $\Delta E\approx \int_{x_1}^{\textrm{Z\textsubscript{R}}}E_{acc} ds$, where $\vb{E}_{acc}$ is the accelerating field and $ds$ is the length element along the trajectory. There are three main factors to consider: the strength of the accelerating field, $E_{acc} \propto E_0 \propto 1/f\#$; the longitudinal distance over which this field is sustained, $Z_\text{R} \propto (f\#)^2$; and dephasing, which controls how much of this distance actually contributes. In the absence of dephasing the energy $\Delta E \propto E_0 Z_\text{R}\propto f\#$. In other words, the weaker focusing decreases the accelerating field, but increases the acceleration length by a greater factor, such that the final electron energy gain will be higher.
There is nevertheless a limit set by dephasing, i.e., when the copropagating electrons leave the accelerating half of the optical cycle. 

Our results show that f\#1 and f\#3 focusing produce similarly energetic electrons (and moreover that the energy is not higher for f\#3), which indicates the potentially important role of dephasing. Indeed Fig.~\ref{fig:MaxEn} shows for f\#3 that saturation of electron energy occurs before one Rayleigh length is reached. 
Dephasing may furher be seen in the spatial distribution of the electron bunches, shown in Fig.~\ref{fig:dephasing}
Whereas for f\#1 the most energetic bunches (B and C) are still in the same half cycle after one Rayleigh length, for f\#3 these bunches have fallen behind by $\lambda_0/4$, arriving in the following half cycle, before even a single Rayleigh length has been reached. This indicates that continuous energy gain has terminated and dephasing has set in.

We should note that our modeling uses the paraxial approximation for the focusing fields \cite{Salamin.2002, popov2008electron}. Furthermore, the charge, energy and angular distribution of individual bunches depends on the CEP which is fixed in simulations and averaged in experiments. While these limitations play a role, as indicated by the differences between results of simulations and experiments, our modelling appears to be sufficient to explain the counter-intuitive phenomena we have observed.

In conclusion, we have investigated VLA experimentally and numerically, using nanotips as electron injectors in different focusing geometries. Our experimental results reveal unexpectedly that VLA does not necessarily benefit from tighter focusing, even though it produces stronger electromagnetic fields, and these are generally considered to be superior for laser-based acceleration schemes. Instead, it is the interplay between diffraction and dephasing that is a key factor.
This is shown by our comparison of f\#1 and f\#3 focusing, which yielded comparable total charges and peak energies.
The decreased laser electric field in the latter case is more than compensated by the increased acceleration length, which we find to be limited mainly to the Rayleigh length.
Our simulations show that dephasing plays a significant role for f\#3 focusing, which points towards possible optimization criteria. We conjecture that in our case an optimum is reached at about f\#2, where the balance between acceleration distance and accelerating field results in peak energies of 22-25~MeV (assuming no dephasing). This paves the way towards generation of high charge, high energy, nano-scale electron bunches that could drive, e.g., an attosecond Thomson X-ray source \cite{Horny2021}.

A.D.A. and S.B. contributed equally to this work. The authors thank Roushdey Salh for the technical support. LV, DH, AG and MM acknowledge the support from Vetenskapsrådet (2019-02376). LV acknowledges the support from Vetenskapsrådet (2020-05111), Knut och Alice Wallenbergs Stiftelse (2019.0140), and Kempestiftelserna (SMK21-0017). SB and AG acknowledge the computational resources provided by the National Academic Infrastructure for Super Computing in Sweden (NAISS-2023/5-43) and (NAISS 2023/22-8).





\end{document}